\documentclass[iop, tighten]{emulateapj}
\usepackage{apjfonts}

\pdfoutput=1

\usepackage{color,hyperref}
\definecolor{linkcolor}{rgb}{0,0,0.5}
\hypersetup{colorlinks=true,linkcolor=linkcolor,citecolor=linkcolor,
            filecolor=linkcolor,urlcolor=linkcolor}
\usepackage{url}
\usepackage{amssymb,amsmath}
\usepackage{subfigure}
\usepackage{booktabs}

\usepackage{natbib}
\def\h2{$\rm H_2$}

\def\kms{km~s$^{-1}$}

\newcommand{\msun}{M$_{\odot}$}

\begin{document}

\title{A Hubble Space Telescope Study of the Enigmatic Milky Way Halo Globular Cluster Crater\altaffilmark{*}}

\author{
Daniel R.\ Weisz\altaffilmark{1,9},
Sergey E.\ Koposov\altaffilmark{2},
Andrew E.\ Dolphin\altaffilmark{3},
Vasily Belokurov\altaffilmark{2},
Mark Gieles\altaffilmark{4},
Mario L.\ Mateo\altaffilmark{5},
Edward W.\  Olszewski\altaffilmark{6},
Alison Sills\altaffilmark{7},
Matthew G.\ Walker\altaffilmark{8}
}

\altaffiltext{*}{Based on observations made with the NASA/ESA Hubble Space Telescope, 
obtained at the Space Telescope Science Institute, which is operated by the 
Association of Universities for Research in Astronomy, Inc., under NASA contract NAS 
5-26555. These observations are associated with program \#13746}
\altaffiltext{1}{Astronomy Department, Box 351580, University of Washington, Seattle, WA, USA; 
dweisz@uw.edu}
\altaffiltext{2}{Institute of Astronomy, University of Cambridge, Madingley Road, Cambridge, CB3 0HA, UK}
\altaffiltext{3}{Raytheon Company, Tucson, AZ, 85734, USA}
\altaffiltext{4}{Department of Physics, University of Surrey, Guildford GU2 7XH, UK}
\altaffiltext{5}{Department of Astronomy, University of Michigan, 311 West Hall, 1085 S. University Avenue, Ann Arbor, MI 48109, USA}
\altaffiltext{6}{Steward Observatory, The University of Arizona, 933 N. Cherry Avenue., Tucson, AZ 85721, USA}
\altaffiltext{7}{Department of Physics and Astronomy, McMaster University, Hamilton, Ontario L8S 4M1, Canada}
\altaffiltext{8}{McWilliams Center for Cosmology, Department of Physics, Carnegie Mellon University, 5000 Forbes Avenue, Pittsburgh, PA 15213, USA}
\altaffiltext{9}{Hubble Fellow}



\begin{abstract}

We analyze the resolved stellar populations of the faint stellar system, Crater, based on deep optical imaging taken with the Advanced Camera for Surveys aboard the Hubble Space Telescope.  The HST-based color-magnitude diagram (CMD) of Crater extends $\sim$4 magnitudes below the oldest main sequence turnoff, providing excellent leverage on Crater's physical properties. Structurally, we find that Crater has a half-light radius of $\sim$20 pc and shows no evidence for tidal distortions.  We model the CMD of Crater under the assumption of it being a simple stellar population and alternatively by solving for its full star formation history.  In both cases, Crater is well-described by a simple stellar population with an age of $\sim$7.5 Gyr, a metallicity of [M/H]$\sim$$-$1.65, a total stellar mass of M$_{\star}\sim1{\rm e}4$ M$_{\odot}$, a luminosity of M$_{\rm V}\sim -5.3$, located at a distance of d $\sim$ 145 kpc, with modest uncertainties in these properties due to differences in the underlying stellar evolution models.  We argue that the sparse sampling of stars above the turnoff and sub-giant branch are likely to be 1.0-1.4 M$_{\odot}$ binary star systems (blue stragglers) and their evolved descendants, as opposed to intermediate age main sequence stars.  Confusion of these populations highlights a substantial challenge in accurately characterizing sparsely populated stellar systems.  Our analysis shows that Crater is not a dwarf galaxy, but instead is an unusually young cluster given its location in the Milky Way's very outer stellar halo. Crater is similar to SMC cluster Lindsay 38, and its position and velocity are in good agreement with observations and models of the Magellanic stream debris, suggesting it may have accreted from the Magellanic Clouds.  However, its age and metallicity are also in agreement with the age-metallicity relationships of lower mass dwarf galaxies such as Leo {\sc I} or Carina.  Despite uncertainty over its progenitor system, Crater appears to have been incorporated into the Galaxy more recently than z$\sim$1 (8 Gyr ago), providing an important new constraint on the accretion history of the Milky Way.
\end{abstract}


\section{Introduction}
\label{sec:intro}

The \textit{Sloan Digital Sky Survey} \citep{york2000} has revolutionized our understanding of the faintest stellar systems.  The deep, wide-field imaging of SDSS facilitated the discovery and characterization of dozens of faint dwarf galaxies and GCs in and around the Milky Way \citep[MW; e.g.,][]{willman2005, willman2005b, belokurov2006, belokurov2007, belokurov2009, belokurov2010, irwin2007, koposov2007, zucker2006, zucker2006b, kim2015}, and has proven transformative for our understanding of the nature of dark matter, the impact of cosmic reionization in the local universe, and how stars form in extremely shallow gravitational potentials \citep[e.g.,][]{simon2007, bovill2009, brown2014, weisz2014a}.

The faint object renaissance catalyzed by SDSS has continued to grow as new wide-field photometric surveys scan previously under-explored regions of the sky.  Within the past year, dozens of new faint objects have been discovered in the south \citep[e.g.,][]{koposov2015, laevens2015, laevens2015b, martin2015, des2015, des2015b}, dramatically increasing the census of known faint stellar systems, including the putative satellite galaxies of the LMC \citep[e.g.,][]{koposov2015b, simon2015, walker2015}.

Among the first objects discovered in this new era was Crater\footnote{Independent co-discovery resulted in multiple names for this object: Crater, Laevens {\sc I}, PSO J174.0675-10.87774.  We have adopted `Crater' for this paper.} \citep{belokurov2014, laevens2014}.  Crater appears to be a predominantly ancient and metal-poor system, similar to the majority of MW globular clusters (GCs) and many of the faintest MW satellites.  However, the presence of stars near the `blue loop' and above the oldest sub-giant branch (SGB) led \citet{belokurov2014} to speculate that Crater may have had multiple generations of star formation, unlike the majority of stellar systems of similar size, luminosity, and proximity to the MW.  The presence of multiple, recent epochs of star formation in Crater would provide qualitatively new insight into how such extremely low-mass objects can retain or accrete fresh gas and form stars, despite having such shallow potentials and being well within the virial radius of the MW.  

However, there has been considerable debate over whether Crater is a GC or a faint galaxy.  Stellar spectroscopy by \citet{kirby2015} demonstrated that three of the four luminous putative blue loop stars are actually low-mass MW foreground stars, effectively ruling out star formation in Crater within last Gyr. This study also revealed a small spread in metallicity and a stellar velocity dispersion that appears consistent with a system made entirely of baryons. They conclude that Crater is a GC.  

In contrast, \citet{bonifacio2015} argue that Crater is more likely to be a dwarf galaxy.  Based on spectroscopy of two RGB stars, they find a velocity dispersion that is larger than expected if only baryons were present. Face value interpretation of this result implies the existence of dark matter and thus categorizes Crater as a galaxy \citep{willman2012}, although \citet{bonifacio2015} acknowledge that the small number of stars and uncertainties on their velocities, make this a tentative conclusion. Moreover, \citet{bonifacio2015} show that the sparse sampling of stars above the oldest SGB are consistent with a $\sim$ 2 Gyr stellar isochrone, which is incompatible with Crater being a simple stellar population (SSP).  
 
However, this region of the color-magnitude diagram (CMD) is also occupied by blue stragglers, the products of binary star evolution, and their descendants, which can mimic the presence of intermediate age single stars.  Unfortunately, the faintness of these sources makes follow-up spectroscopy prohibitively expensive at this time and we must rely on other means for interpretation.  

In this paper, we present deep optical imaging of Crater taken with the \emph{Hubble Space Telescope} (HST) and characterize its stellar populations by analyzing the resulting deep CMD.  The HST-based CMD extends several magnitudes fainter than existing ground-based data, providing new perspective on the nature of Crater.  Using CMD analysis methods that are routinely applied to Local Group and nearby dwarf galaxies \citep[e.g.,][]{weisz2011a, weisz2014a}, we undertake a detailed characterization of Crater's stellar populations and conclude that it is a GC and not a dwarf galaxy.  

This paper is organized as follows.  In \S \ref{sec:obs} and \ref{sec:data}, we present the observations, describe the data reduction, and discuss the CMD and derive the structural parameters of Crater.  In \S \ref{sec:match}, we summarize our method of CMD analysis, and we present the results in \S \ref{sec:results}.  Finally, in \S \ref{sec:discussion}, we examine Crater in the context of known dwarf galaxies and the MW GC population and discuss possible formation and accretion scenarios.

\begin{figure}
\epsscale{0.9}
\plotone{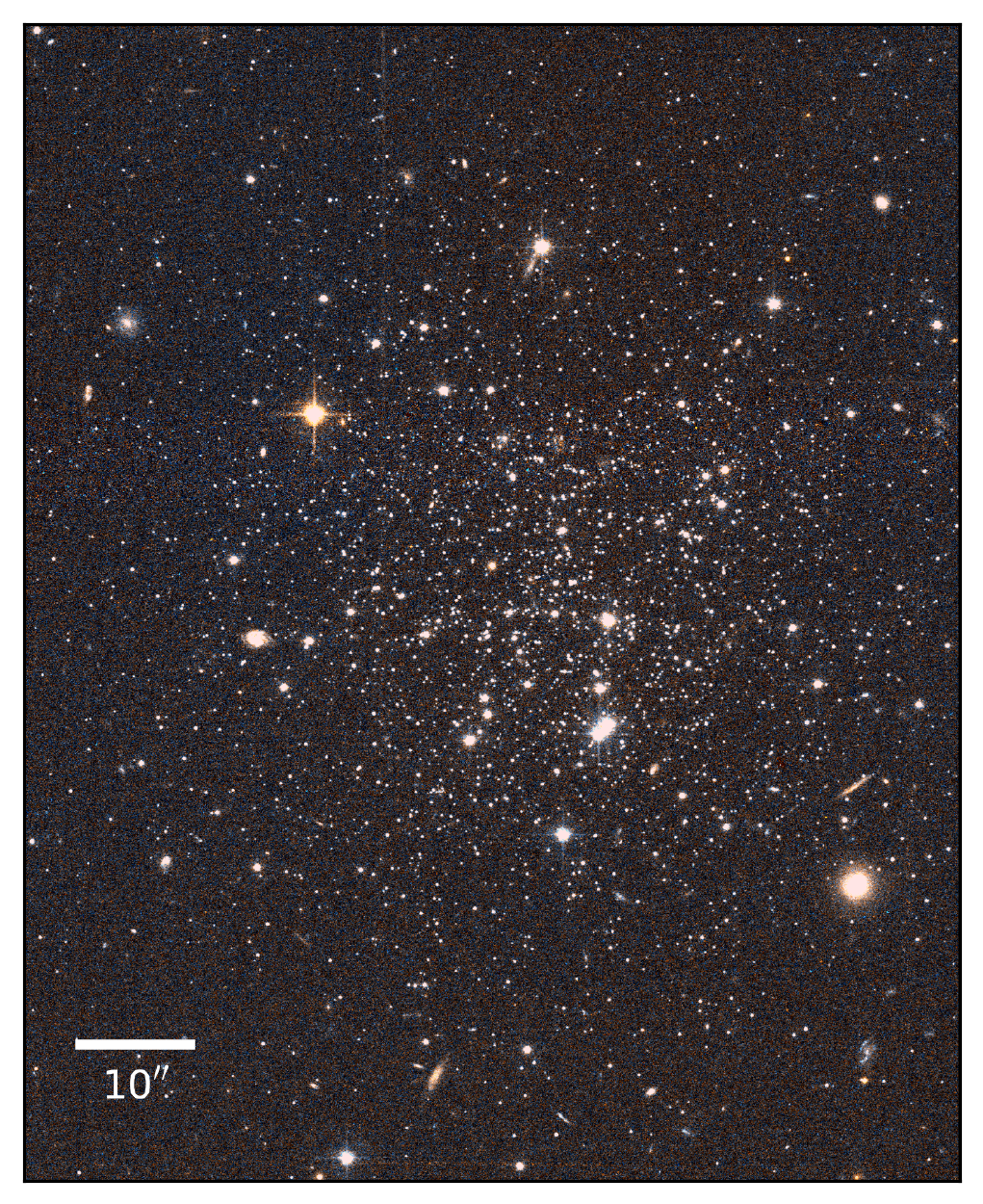}
\caption{A colorized cutout of the composite F606W and F814W HST/ACS image of Crater.}
\label{fig:crater_image}
\end{figure}

\section{Observations \& Data Reduction}
\label{sec:obs}

Observations of Crater were taken with Advanced Camera for Surveys \citep[ACS;][]{ford1998} aboard HST on November 11-12, 2014 as part of HST-GO-13746 (PI: M.\ Walker).  The observations consisted of deep integrations in F606W ($R$-band) and F814W ($I$-band) with multiple exposures to mitigate the impact of cosmic rays. We did not dither to fill the chip gap as Crater easily fit on one ACS chip.  The basic  properties of Crater and our observations are listed in Table \ref{tab:global} and a false color image of Crater is shown in Figure \ref{fig:crater_image}.

We performed point spread function photometry on each of the charge transfer efficiency corrected (\texttt{flc}) images using \texttt{DOLPHOT}, an updated version of \texttt{HSTPHOT} \citep{dolphin2000b} with HST-specific modules.  The parameters used for our photometry follow the recommendations in \citet{williams2014}.  

We culled the catalog of detected objects to include only well-measured stars by requiring: SNR$_{\mathrm{F606W}} > 5$, SNR$_{\mathrm{F814W}} > 5$, (sharp$_{\mathrm{F606W}}$ $+$ sharp$_{\mathrm{F814W} }$)$^2 < 0.1$, and (crowd$_{\mathrm{F606W}}$ $+$ crowd$_{\mathrm{F814W} }$)$ < 1.0$.  Definitions of each of these parameters can be found in \citet{dolphin2000b}.  We characterized completeness and photometric uncertainties using $\sim$50,000 artificial star tests (ASTs).  Our HST photometric catalog is available through MAST\footnote{\url{https://archive.stsci.edu/hst/}}.

\begin{deluxetable}{lc}
\tablecolumns{2}
\tablehead{
\colhead{Quantity} &
\colhead{Value}
}
\startdata
RA (J2000) &  11:36:16.5 \\
DEC (J2000) & $-$10:52:37.1 \\
Obs. Dates & Nov 11-12 2014 \\
Exp. Time (F606W,F814W) (s) & 3915, 4095 \\
50\% Completeness (F606W,F814W) & 27.6, 27.1   \\
Stars in CMD & 3620 \\
Distance (kpc) & 145 $\pm$ 3 \\
M$_{V}$ & $-$5.3 $\pm$ 0.1  \\
M$_{\star}$ (10$^3$ \msun) & 9.9$^{+0.1}_{-0.05}$  \\
r$_{1/2, \rm{Plummer}}$ (\arcmin) & 0.46 $\pm$ 0.01 \\
r$_{1/2, \rm{Plummer}}$ (pc) & 19.4 $\pm$ 0.4 \\
r$_{1/2, \rm{exponential}}$ (\arcmin) & 0.43 $\pm$ 0.01 \\
1 - b/a & $<$0.055 (90\%) 
\enddata
\tablecomments{Observational and structural properties of Crater.  The distance, stellar mass, absolute luminosity, and structural parameters were computed from our analysis of Crater's HST-based CMD (see \S \ref{sec:results}).}
\label{tab:global}
\end{deluxetable}

\section{The Data}
\label{sec:data}

\subsection{The Color-Magnitude Diagram and Membership Identification}
\label{sec:CMD}

\begin{figure}
\epsscale{1.2}
\plotone{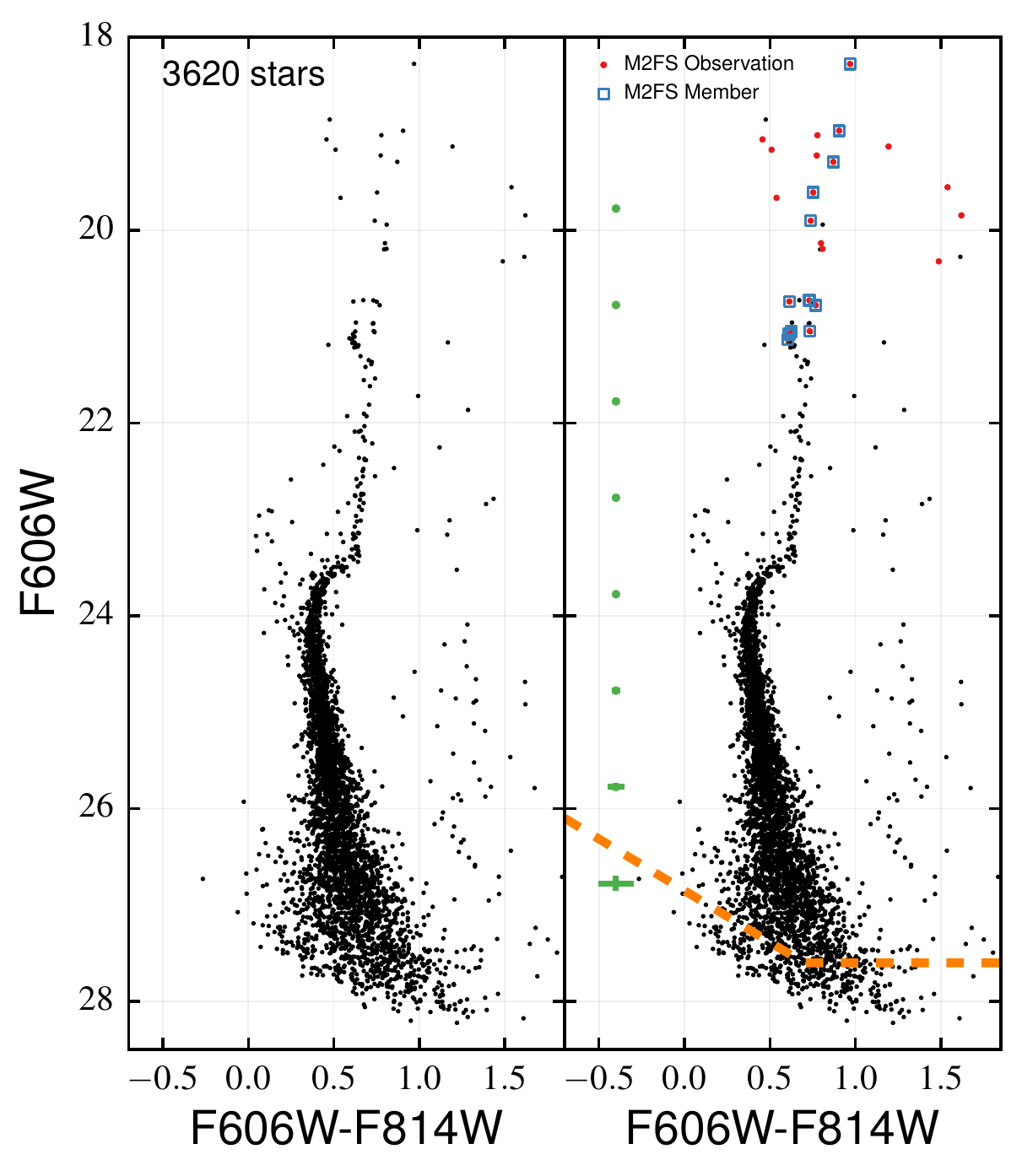}
\caption{The HST/ACS CMD of Crater.  The red points in the right panel indicate stars with Magellan/M2FS spectroscopy.  Those in blue squares are confirmed members of Crater.  The photometric uncertainties are shown in green and 50\% completeness limit, as determined by ASTs, is indicated by the orange dashed line.}
\label{fig:crater_cmd}
\end{figure}

We plot the HST/ACS CMD of Crater in Figure \ref{fig:crater_cmd}. Crater is clearly a predominantly older stellar system ($\gtrsim$3 Gyr), based on the lack of a luminous main sequence (MS).  In terms of population complexity, the narrowness of the RGB, oldest main sequence turnoff (MSTO), and extent of the main SGB suggests that the majority of stars in Crater were formed with a similar age and metallicity.  The presence of a red clump and absence of a blue horizontal branch indicate that Crater is not an ancient and extremely metal-poor system, such as M92 or MW satellites of similar luminosity \citep[e.g.,][]{vandenberg2013, brown2014}. As noted in \citet{belokurov2014}, Crater's horizontal branch appears to be unusually red for its metallicity, when compared to other MW GCs.

Beyond these dominant attributes, the CMD of Crater exhibits several secondary features.  The first is a set of four luminous stars located at F606W$\sim$19 and F606W$-$F814W$\sim$0.5 identified by \citet{belokurov2014} as putative blue loop stars with ages $<$ 1 Gyr.  If these are \textit{bone fide} `blue loop' stars, we would expect to see a larger number of luminous blue stars corresponding to the young main sequence brighter than F606W$\sim$21.  None are observed.  

A second interesting feature is located immediately above the primary SGB at F606W$\sim$23.5 and F606W$-$F814W$\sim$0.2.  The colors and magnitudes of these stars are consistent with either being $\sim$2-5 Gyr old main sequence stars or blue stragglers.  From the bright end of this feature, another set of stars extends diagonally up to F606W$\sim$22, where it intersects with the RGB.  \citet{bonifacio2015} identify these stars as being consistent with a $\sim$ 2 Gyr MSTO and SGB with a photometric metallicity of [M/H] $= -1.5$.  

Finally, the set of stars located to the red of the RGB at F606W$-$F814W$\sim$1.3-1.5 at all magnitudes are likely low-mass MW foreground stars.  Given the small ACS field of view, the relatively large number of foreground stars suggests there is a non-negligible amount of foreground star contamination.  Such stars are also likely to overlap with other parts of the CMD, notably the RGB, which can lead to confused interpretation of the stellar populations.

To aide in identification and removal of non-members, we use stellar spectroscopy obtained with Michigan/Magellan Fiber System \citep[M2FS;][]{mateo2012} on the Magellan/Clay telescope as described in M.\ Mateo (in prep.).  The M2FS observations are indicated by red points in the right panel of Figure \ref{fig:crater_cmd}.  Only those enclosed within blue squares are likely members.  The M2FS spectroscopy shows that three of the four putative blue loop stars are not members of Crater,  Instead, they are foreground stars, which is consistent with the spectroscopic findings presented in \citet{kirby2015}.  The \citet{kirby2015} analysis of the fourth `blue loop' star has proven inconclusive.  It has a systematic velocity that is close to that of Crater, which would favor it being a member.  However, it would take an extremely unusual star formation history (SFH) or initial mass function (IMF) sampling to produce a single, young blue star.

The M2FS spectroscopically confirmed members trace out a narrow RGB and the red clump.  Unfortunately, spectroscopy of stars fainter than the red clump is prohibitively expensive at this time. 

\subsection{Structural Parameters}

We leverage the exquisite depth of the HST data to investigate the spatial structure of Crater.  Specifically, we model the distribution of stars in Crater with two-dimensional elliptical Plummer and exponential models following the procedure described in \citet{koposov2015}. 

\begin{figure}
\epsscale{1.2}
\plotone{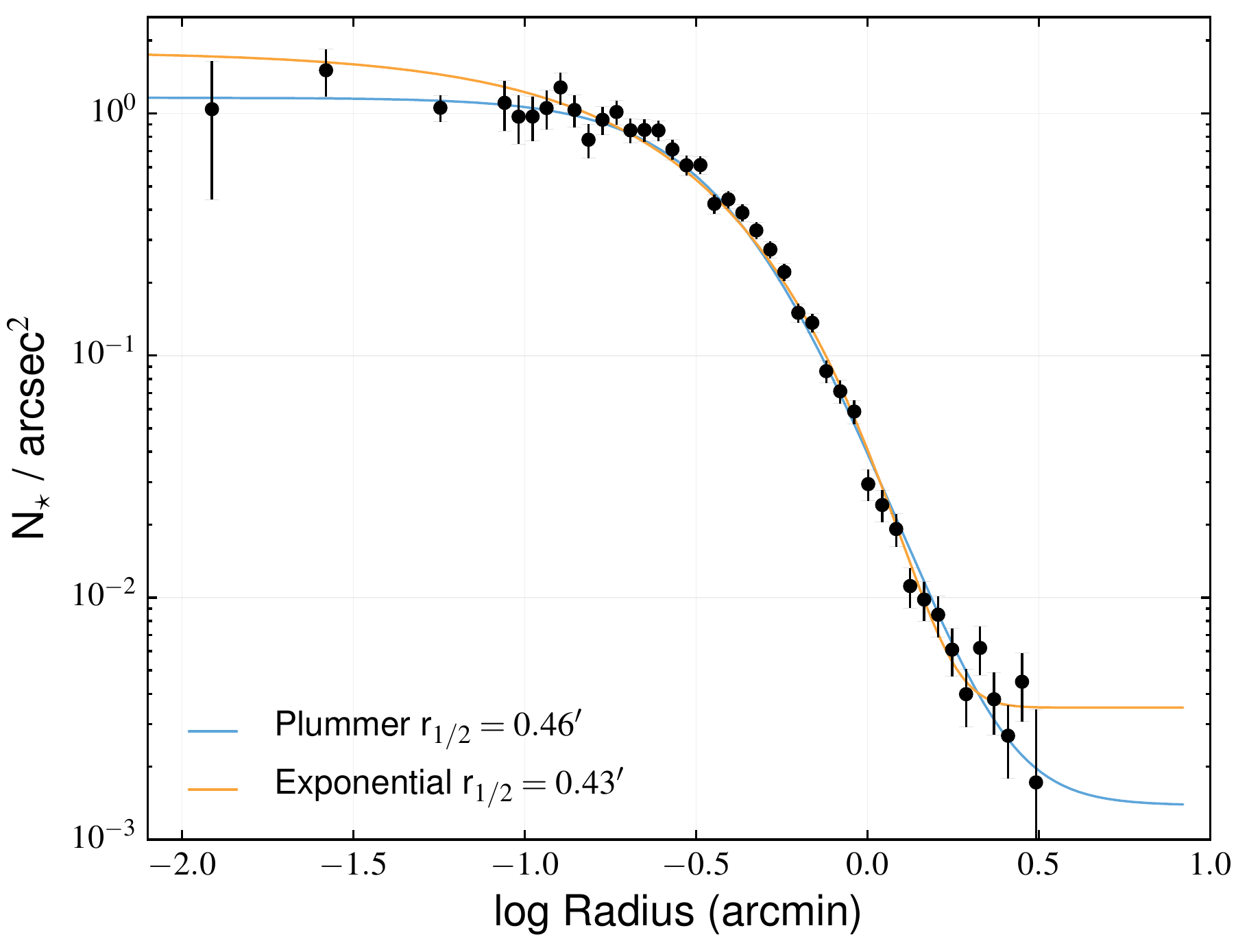}
\caption{The radial stellar density profile of Crater based on our HST photometric catalog.  The red and blue lines indicate the best fit Plummer and exponential profiles, respectively.}
\label{fig:crater_sb}
\end{figure}

For the purposes of this analysis, we only use objects classified as stars with F606W$<$27 to ensure high completeness across the entire catalog.  We also remove obvious contamination from MW dwarf stars red-ward of the Crater main sequence and RGB (F606W$-$F814W $ \gtrsim 1$). 

The model parameters we use to describe Crater's structure are: Crater's center, half-light radius, ellipticity, positional angle, surface density of background stars, and central surface density of Crater. For each model (Plummer, exponential), we sample the posterior probability distribution with a Markov chain Monte Carlo (MCMC) technique, assuming uniform priors on all parameters.  The resulting marginalized distributions for all parameters are listed in Table \ref{tab:global}.  Aside from ellipticity, all of the marginalized distributions are well-described by a Gaussian.  In Figure \ref{fig:crater_sb}, we plot the one-dimensional density profile of Crater together with the most probable Plummer and exponential models.  The observed density profile shows no substantial deviations from a Plummer model.

From this modeling, we find that Crater has an ellipticity of $<$ 0.055 at 90\% confidence level, indicating it is consistent with being circular. The measured half-light radius from the Plummer fit is $r_{1/2} = 0\arcmin.46\pm0.01$, which is consistent with measurement of \citet{laevens2014} and is slightly smaller than the estimate in \citet{belokurov2014}. The half-light radius measured using an exponential density profile is $r_{1/2} = 0\arcmin.43\pm0.01$.  Assuming a heliocentric distance of 145 kpc, the half-light radius of Crater is $19.4 \pm 0.4\,$pc.  The combination of being circular and well-described by a Plummer profile suggests that Crater is unlikely to have experienced drastic tidal stripping.

\section{Methodology}
\label{sec:match}

We analyze the stellar populations of Crater using the CMD modeling software package \texttt{MATCH} \citep{dolphin2002}.  In brief, \texttt{MATCH} requires a user specified stellar evolution library, stellar initial mass function (IMF), binary fraction, and search ranges in distance, extinction, age, and metallicity.  For a given combination of these parameters, \texttt{MATCH} constructs a set of SSPs that are linearly combined to form a composite CMD. The weight of each SSP is the star formation rate (or total stellar mass) at that age/metallicity combination.  The composite CMD is then convolved with the observational noise model (photometric uncertainties and completeness) as determined by the ASTs. Finally, the synthetic and observed CMDs are compared in bins of color and magnitude of specified size (0.05 and 0.1 mag, respectively), and the probability of the observed CMD given the synthetic CMD is computed using a Poisson likelihood function.   More details on the general methodology of \texttt{MATCH} can be found in \citet{dolphin2002}.

\begin{deluxetable}{lcc}
\tablecolumns{3}
\tablehead{
\colhead{Quantity} &
\colhead{Range} &
\colhead{Resolution} 
}
\startdata
$(m-M)_0$ & 20.60-21.10 & 0.05, 0.01 \\
$\mathrm{A_{V}}$ & 0.0-0.5 & 0.05, 0.01 \\		
IMF & \citet{kroupa2001} & fixed \\
log(Age) & 9.0-10.15 & 0.05 \\
$[M/H]$  & $-2.3$ to $-0.5$ & 0.05 \\
$[\alpha/Fe]$  & $-0.2, 0.0, +0.2, +0.4$ & fixed\\
binary fraction & 0.1, 0.35, 0.6 & fixed 
\enddata
\tablecomments{Parameters and their ranges and resolutions used as input into \texttt{MATCH}. Parameters with multiple resolutions indicated were solved for iteratively: first through a large search with the coarser resolution and then via a focused search at the higher resolution.  $\alpha$-enhancements are only currently available for the Dartmouth models.  We tested several values for the binary fraction, but found find that it did not substantially affect determination of the other physical parameters.}
\label{tab:paramvals}
\end{deluxetable}

We model Crater's CMD in two ways: first assuming it is an SSP and second by solving for its full SFH.  In both cases we used the parameters listed in Table \ref{tab:paramvals} and employed two different stellar evolution libraries, Dartmouth \citep{dotter2008} and PARSEC \citep{bressan2012}, in order to quantify the sensitivity of our result to the choice of stellar evolution model.  While the PARSEC models are currently only available with solar-scaled abundances, i.e., [$\alpha$/Fe]$=0.0$, the Dartmouth models allowed us to explore several $\alpha$-enhancements ranging from [$\alpha$/Fe]$=-0.2$ to $+0.4$.  

We took an iterative approach to modeling the CMD such that we started with coarse resolution searches in distance and extinction (0.05 dex resolution) and, upon convergence, used a finer grid resolution (0.01 dex) for the final solutions.  

\begin{figure}[t!]
\epsscale{1.2}
\plotone{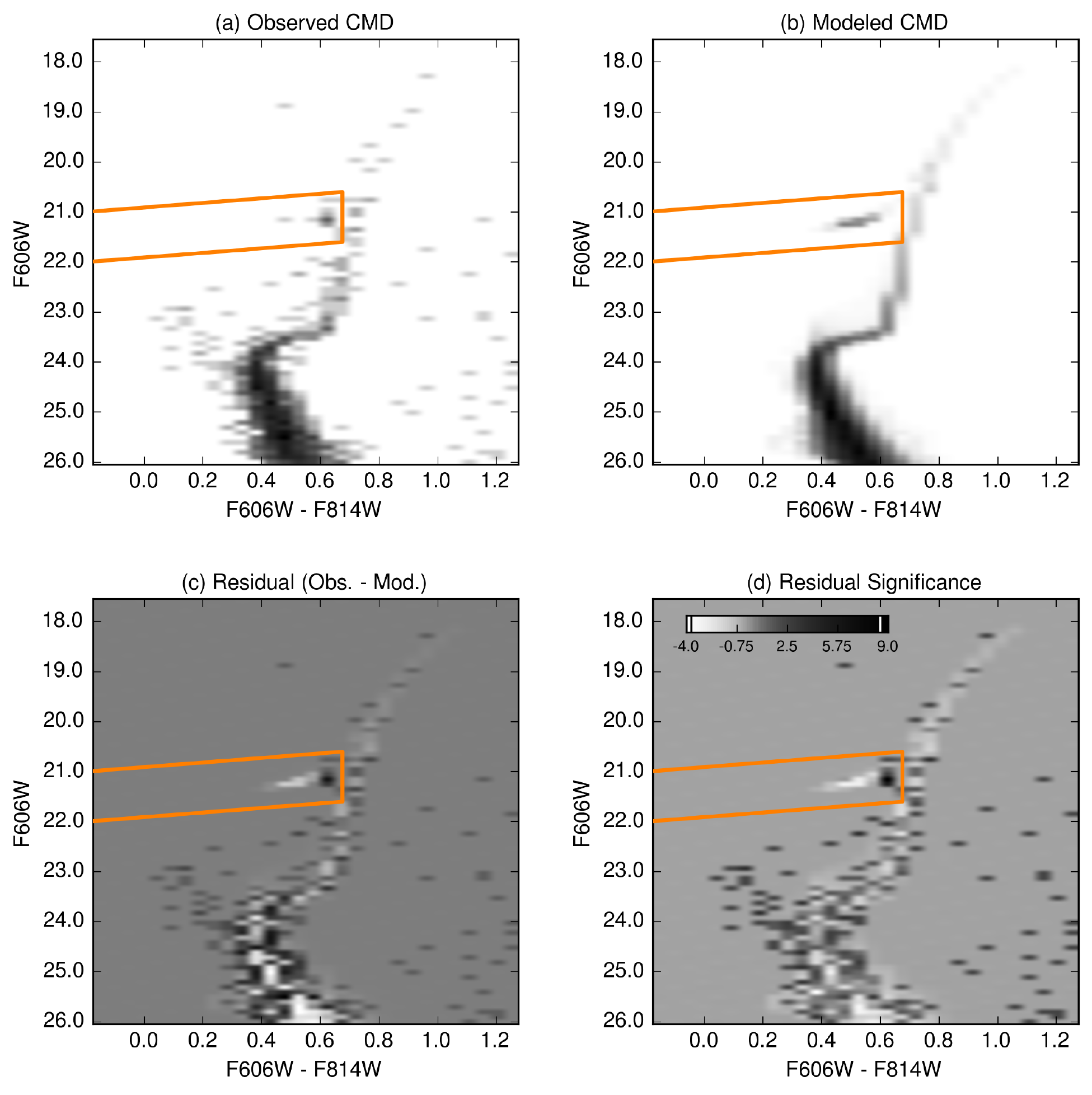}
\caption{Hess diagrams illustrating our SSP model of the Crater CMD using the Dartmouth models with [$\alpha$/Fe]$= +0.4$.  The area inside the orange region was excluded from the fitting.  Panels (a) and (b) show the observed and most likely model CMDs.  Panels (c) and (d) show the residual and residual significance diagrams.  The color bar in panel (d) is in units of $\sigma$, where white and black represent the most extreme deviations.  Overall, the Crater is well fit by an SSP.  The physical parameters derived from this fit are listed in Table \ref{tab:derivedparams}.}
\label{fig:dartssp}
\end{figure}

Following several previous analyses of deep HST-based CMDs \citep[e.g.,][]{weisz2012b, weisz2014m31}, we masked out the red clump and horizontal branch regions of the observed CMD for the fitting process, as indicated in Figure \ref{fig:dartssp}.  The physics of these evolutionary phases are highly uncertain \citep[e.g.,][]{gallart2005}, and their inclusion in CMD fitting process can be problematic. Instead, we rely on the more secure MSTO and sub-giant sequences for this analysis.  

To model the possible contamination from the MW for stars fainter than where M2FS spectroscopy is available (i.e., m$_{F606W} >$ 21), we use a statistical model of the MW foreground based on analysis presented in \citet{dejong2010b}.

We determine uncertainties in Crater's properties differently for the SSP and complex stellar population assumptions.  When we require Crater to be an SSP, we compute likelihood values over every possible combination of parameters in the grid.  This approach has the advantage of sampling the entirety of likelihood space, which allows for convenient marginalization and full consideration of stochastic effects \citep[e.g.,][]{weisz2015b}.  Because of the smoothness of the likelihood surface, we marginalize over a finely interpolated grid to measure the most likely values and associated confidence intervals for each parameter to a higher degree of precision than is afforded by the native resolution.

\begin{deluxetable*}{lccccc}
\tablecolumns{6}
\tablehead{
\colhead{Property} &
\colhead{Dartmouth} &
\colhead{Dartmouth} &
\colhead{Dartmouth} &
\colhead{Dartmouth} &
\colhead{PARSEC} \\
\colhead{} &
\colhead{[$\alpha$/Fe] $= -0.2$} &
\colhead{[$\alpha$/Fe] $= 0.0$} &
\colhead{[$\alpha$/Fe] $= +0.2$} &
\colhead{[$\alpha$/Fe] $= +0.4$} &
\colhead{[$\alpha$/Fe] $= 0.0$} 
}
\startdata
Age (Gyr)			& 6.7$\pm$0.4  & 7.5$\pm$0.4  & 7.5$\pm$0.4 	& 7.5$\pm$0.4 	& 6.7$\pm$0.4\\
$[M/H]$ 			& $-$1.03$\pm$0.02 & $-$1.33$\pm$0.03 & $-$1.55$\pm$0.04 & $-$1.66$\pm$0.04 & $-$1.47$\pm$0.03 \\
$\mathrm(m-M)_{0}$ & 20.83$\pm$0.03  & 20.82$\pm$0.03  & 20.82$\pm$0.03  & 20.81$\pm$0.03 	& 20.82$\pm$0.03\\
$\mathrm{A_{V}}$ 	& 0.09$\pm$0.03 & 0.10$\pm$0.03 & 0.10$\pm$0.03 & 0.10$\pm$0.03	& 0.11$\pm$0.03 \\
Mass (10$^3$ \msun) & 9.7$^{+0.1}_{-0.05}$ & 9.8$^{+0.1}_{-0.05}$ & 9.8$^{+0.1}_{-0.05}$ & 9.9$^{+0.1}_{-0.05}$ & 10.0$^{+0.1}_{-0.07}$
\enddata
\tablecomments{Derived properties for Crater as an SSP.  We list the most likely value and the 68\% confidence intervals. The uncertainties are only statistical in nature, i.e., they scale with the number of stars on the CMD.}
\label{tab:derivedparams}
\end{deluxetable*}

When solving for the full SFH, computing a full grid of solutions is not tractable in a reasonable amount of computational time.  Instead, \texttt{MATCH} finds the most likely solution, and uses an MCMC routine to explore the likelihood surface around it.  Specifically, we use a Hamilonian Monte Carlo \citep{duane1987} approach to sampling SFH space as described in \citet{dolphin2013}. For quantifying the uncertainties on Crater's most likely SFH, we used 5,000 MCMC realizations.

\section{Results}
\label{sec:results}

\subsection{Crater as a Simple Stellar Population}
\label{sec:craterssp}

As shown in Figure \ref{fig:dartssp}, Crater is well-described by an SSP.  Qualitatively, the observed and model CMDs have similar appearances and the MSTO, SGB, and RGB have luminosities, colors, and stellar densities that appear in excellent agreement.  This impression is quantitatively reinforced by the residual significance diagram (panel (d)). This diagram shows that the model CMD does an excellent job of reproducing the observed CMDs;  there are no systematic mismatches between the model and observed CMDs (e.g., clumps or streaks of all black or white bins), which indicates good data-model agreement.

However, there are two discrepant areas that warrant discussion.  First, the luminous blue stars above the predominant MSTO are an area of mismatch.  In the event that these are intermediate age MS stars, our model of an SSP would not be appropriate for Crater, leading to this type of data-model disagreement.  On the other hand, these stars may be blue stragglers, which are not included in the PARSEC or Dartmouth libraries, again resulting in a poor data-model match.  We discuss the nature of these stars below in \S \ref{sec:cratersfh}.  

The second poorly fit region is the red clump.  As discussed in \S \ref{sec:match}, the observed RC was masked from the fit.  Given its exclusion, and outstanding issues in the physics of modeling the red clump \citep[e.g.,][]{gallart2005}, it is not surprising that this feature is not well-described by the models.  However, because we have excluded it from the fit it does not affect our characterization of Crater.

\begin{figure}[t!]
\epsscale{1.2}
\plotone{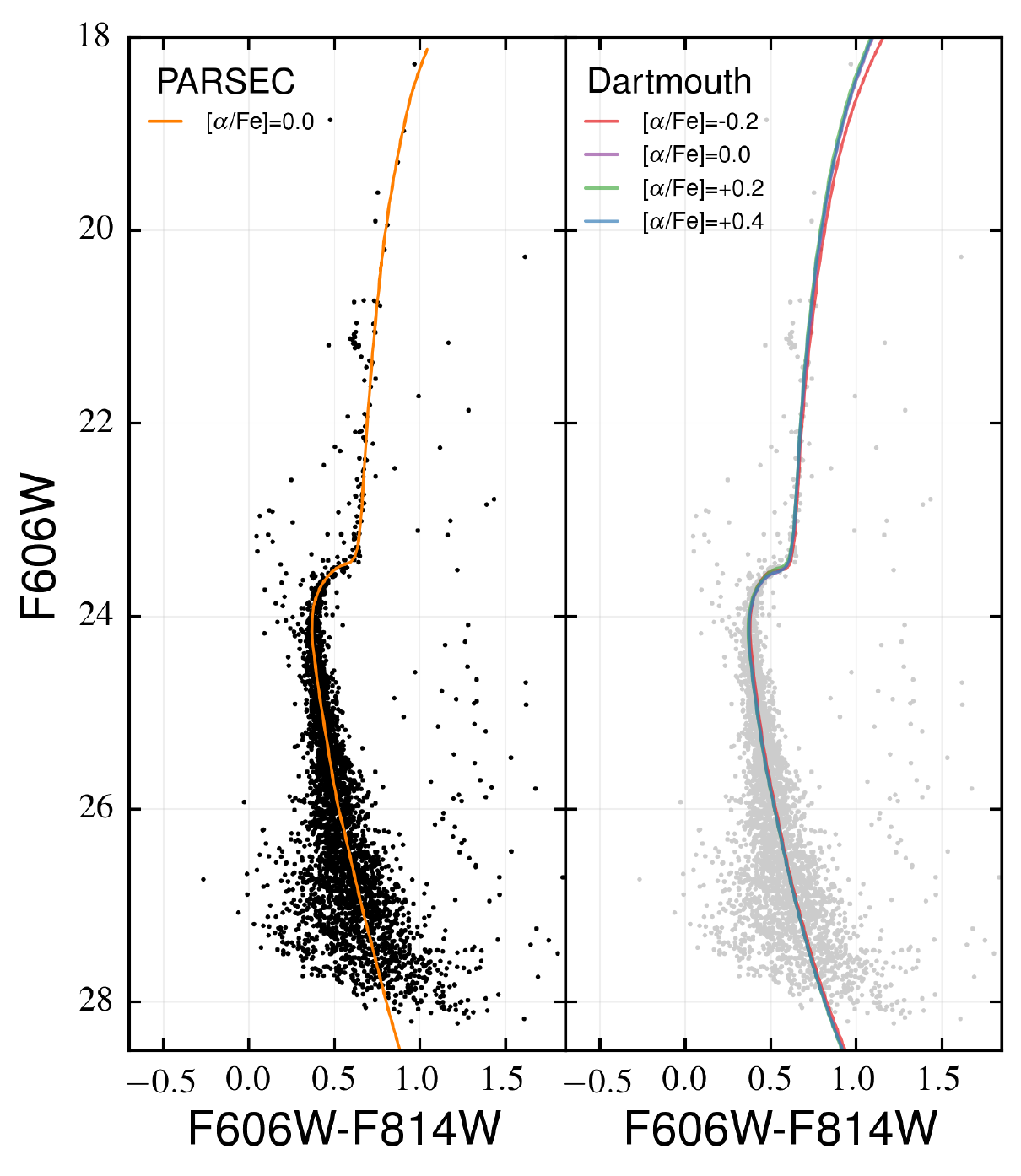}
\caption{CMDs of Crater with the best fitting PARSEC and Dartmouth isochrones over-plotted.   Variations in the inferred physical properties are due to subtle differences in the model shapes of the oldest MSTO, SGB, and slope of the RGB.  Point in the right-hand panel have been greyed out to provide increased visual contrast with the isochrones.}
\label{fig:isocmd}
\end{figure}

The derived physical parameters of Crater are listed in Table \ref{tab:derivedparams}.  In general, the different models produce compatible values for distance, extinction, and total stellar mass.  The most notable variations are in age and metallicity.  Within the Dartmouth models, the age and metallicity are sensitive to the level of $\alpha$-enhancement.  While the age only exhibits modest variations as a function of [$\alpha$/Fe], the mean metallicity varies by $\sim$ 0.6 dex.  Of these solutions, the observed CMD is best described by an [$\alpha$/Fe]$= + 0.4$, although values of 0.0 and $+$0.2 cannot be ruled out at a confidence level $>$95\%.  The model with [$\alpha$/Fe]$= -0.2$ provides a drastically worse fit to the CMD.  In addition to being a marginally better fit, the model with [$\alpha$/Fe]$= + 0.4$ also produces a metallicity that is in good agreement with spectroscopic measurements.  However, as shown in Figure \ref{fig:isocmd}, the color and magnitude differences in isochrones with various amount of $\alpha$-enhancements are quite subtle.  

To examine the sensitivity of Crater's parameters to stellar physics, we compare derived parameters for the solar-scaled Dartmouth and PARSEC models (see Figure \ref{fig:isocmd}).  These two libraries show an age difference of $\sim$0.8 Gyr and a metallicity difference of 0.15 dex.  The age difference is due to slight variations in the shape of the MSTO and SGB between the two models, as shown in Figure \ref{fig:isocmd}.  The amplitude of this difference reflects the uncertainties due to choices in underlying stellar physics and is inline with the expected precision for absolute ages of stars and SSPs, which is $\gtrsim$10\% of the age of the object \citep[e.g.,][]{soderblom2010, cassisi2014}.  The $\sim$0.15 dex offset metallicity is due to subtle differences in the RGB slopes of the two models.

Finally, we computed the stellar mass and integrated luminosity for Crater by summing up the mass and light from the best fit SSP models. While this approach does not include the contribution of blue stragglers, it does mitigate the contribution of non-member stars to the total luminosity.  The stellar mass measurements for Crater are listed in Table \ref{tab:derivedparams}, and the integrated V-band magnitude, from the Dartmouth [$\alpha$/Fe]$= + 0.4$  model, is listed in Table \ref{tab:global}.  We calculated uncertainties on these quantities using the SSP solutions that fell within the 68\% confidence intervals of the best fit models.  Our total integrated luminosity for Crater is similar to that presented in \citet{belokurov2014} and \citet{laevens2014}, and variations of total luminosity between the stellar models are negligibly small.

\subsection{Crater as a Complex Stellar Population}
\label{sec:cratersfh}

\begin{figure}[t!]
\epsscale{1.2}
\plotone{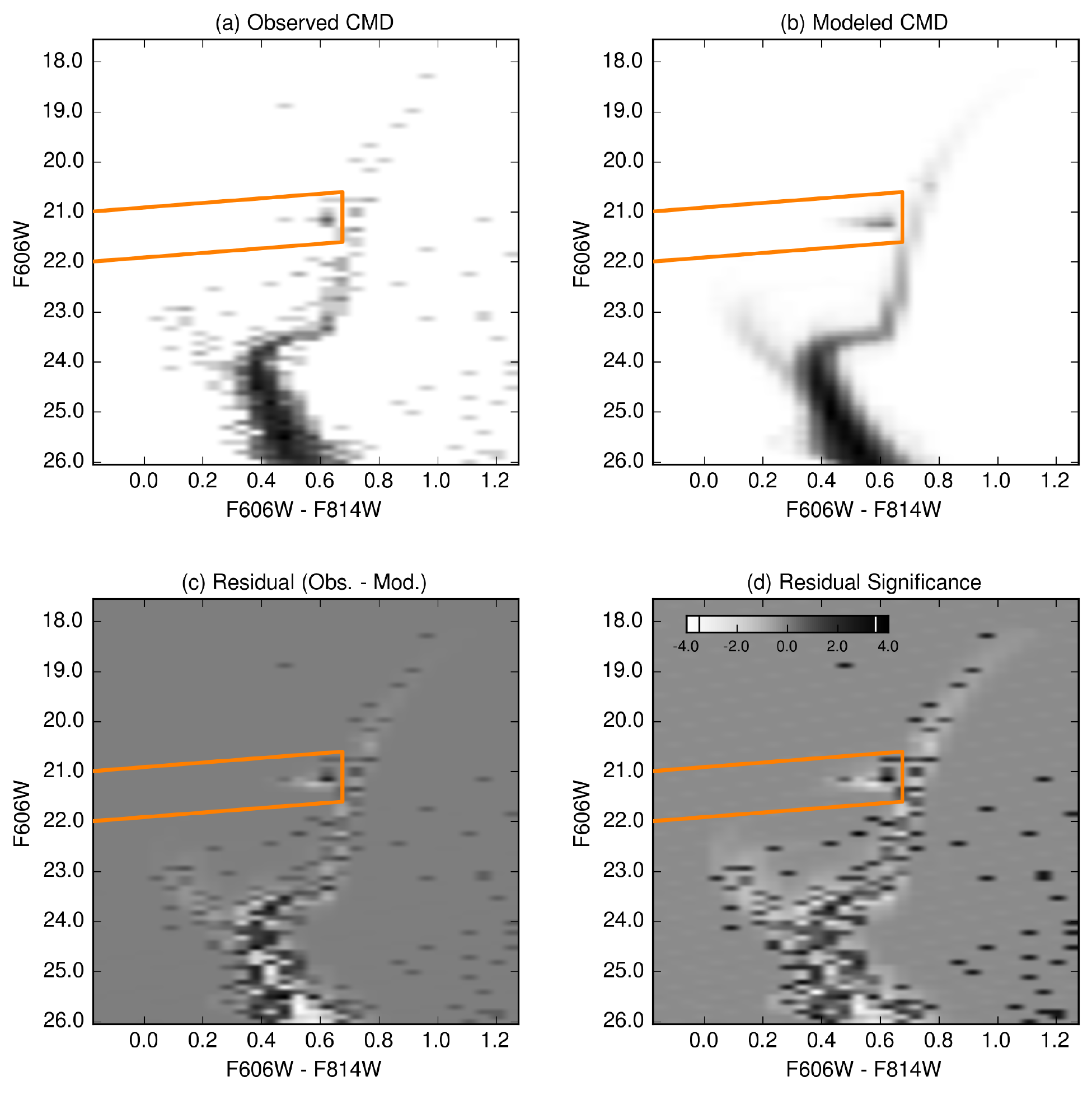}
\caption{Same as Figure \ref{fig:dartssp}, but for Crater's full SFH.}
\label{fig:parsecsfhcmd}
\end{figure}

\begin{figure}[t!]
\epsscale{1.2}
\plotone{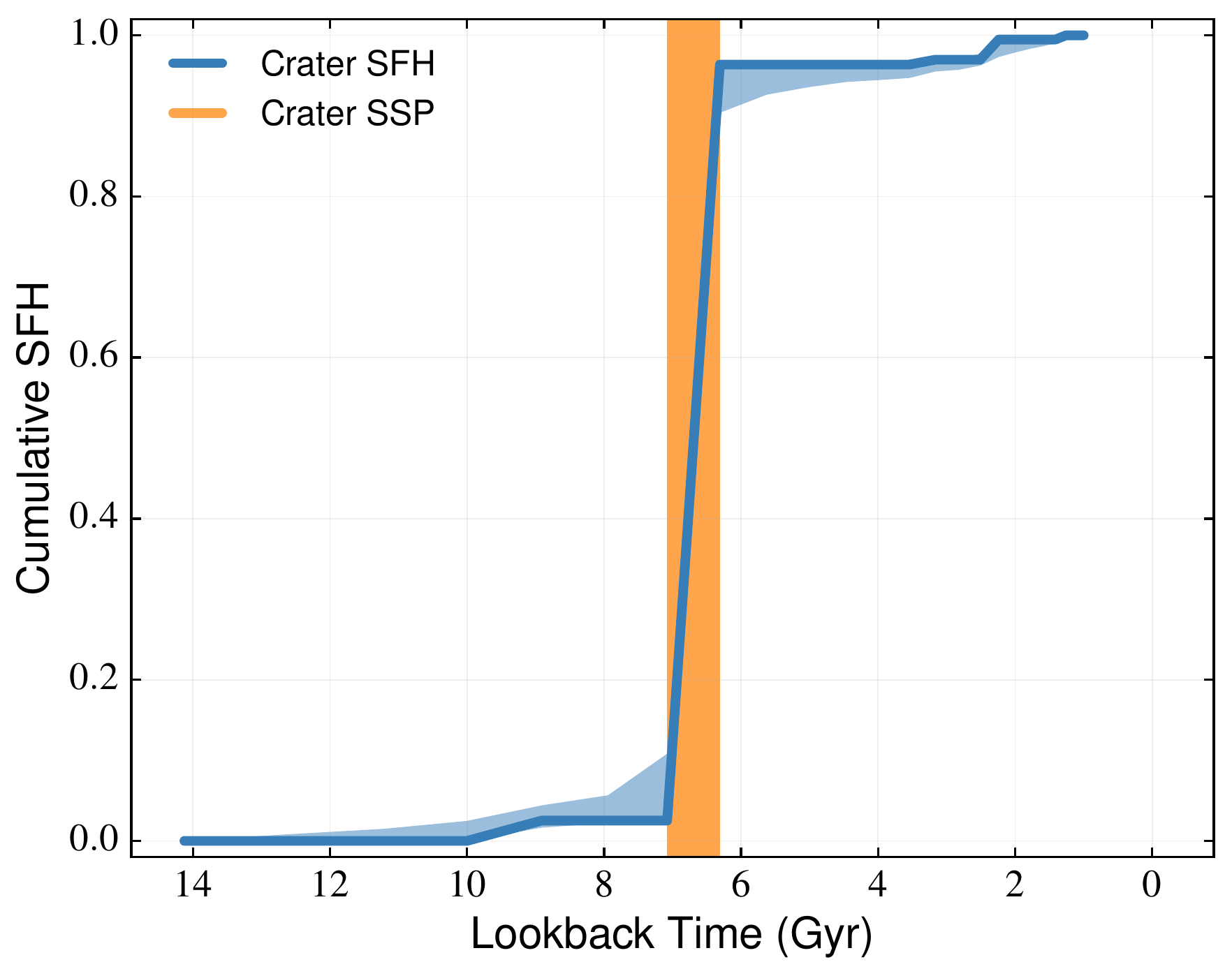}
\caption{A comparison between the full SFH of Crater (blue) and its best fit SSP (orange) as measured with the Dartmouth models ([$\alpha$/Fe]$=+0.4$).  The shaded blue region indicates the 68\% confidence interval in the SFH as determined with 5000 MCMC iterations.  The full SFH shows that $>$95\% of Crater's total stellar mass formed at a single age, which is consistent with it being an SSP.  The small percentage Crater's stellar mass that formed $\sim$ 3 Gyr ago is due to the code modeling likely blue stragglers as intermediate age main sequence stars, and is unlikely to be genuine intermediate age star formation.}
\label{fig:cratersfh}
\end{figure}

We now relax the SSP assumption and model the full SFH of Crater, i.e., we allow it to be fit by an arbitrary sum of SSPs as described in \S \ref{sec:match}.  In Figure \ref{fig:parsecsfhcmd}, we plot the model CMD for the full SFH fit and in Figure \ref{fig:cratersfh} we show the cumulative SFH (blue; the fraction of stellar mass formed prior to a given epoch) and the best fitting Dartmouth SSP with [$\alpha$/Fe]$= +0.4$ (orange). For simplicity, we only show the results for a single Dartmouth model, and note that the fitting with other Dartmouth or PARSEC models produce a similar result.  

Compared to the SSP scenario, Figure \ref{fig:parsecsfhcmd} shows that there are fewer highly discrepant regions in the residual significance CMD (panel (d)).  This is due to the increased the number of free parameters in the model.  The only poorly modeled region is the MW foreground population. The blue stars above the MSTO have been modeled by populations of intermediate age main sequence stars.

The main result of this analysis in shown in Figure \ref{fig:cratersfh}.  The full SFH shows that $>$95\% of the stellar mass in Crater formed in a single event around $\sim$7.5 Gyr ago.  This is fully consistent with the best fit SSP age as discussed in \S \ref{sec:craterssp}, and reinforces that Crater is well-described by an SSP, even when a complex population is allowed.

The remaining $\sim$5\% of the stellar mass formed either slightly before or after the main epoch.  The small amount of mass formed prior to $\sim$7.5 Gyr ago can be attributed to the code compensating for slight mismatches between the isochrone and observed CMD.  Similarly, the small amount of mass formed $\sim$ 3 Gyr ago is the result of fitting the luminous blue stars as main sequence stars.

\begin{figure}[t!]
\epsscale{1.2}
\plotone{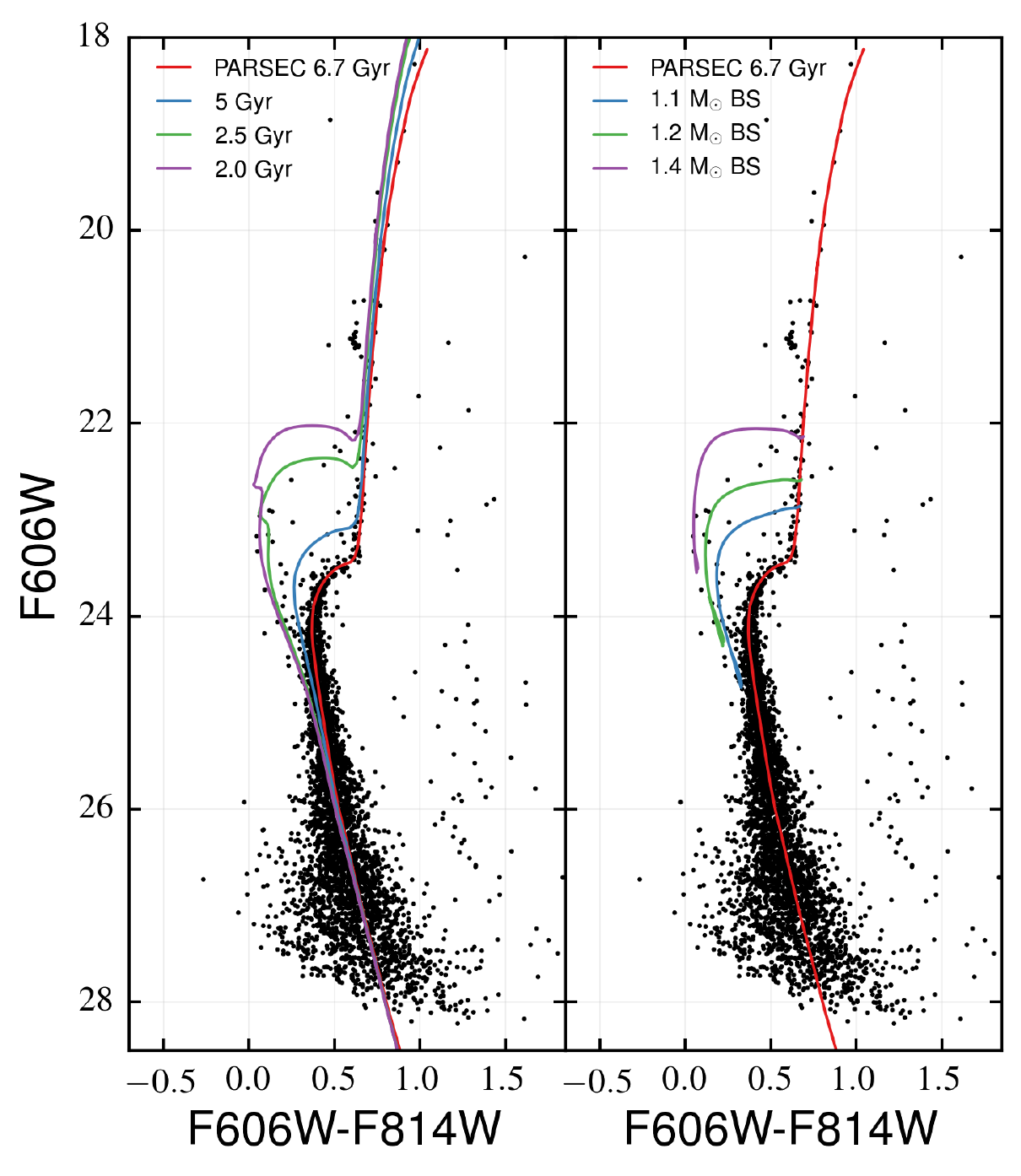}
\caption{CMDs of Crater with select PARSEC isochones over-plotted in the left panel, and tracks of collisional blue stragglers from \citet{sills2009} over-plotted on the right.  A comparison of the two set of models illustrates the similarity between the two intermediate age single stars and blue stragglers.}
\label{fig:bss_cmd}
\end{figure}

However, as illustrated in Figure \ref{fig:bss_cmd}, it is equally plausible that these blue stars are blue stragglers.  The left-hand panel shows that while these single-star isochrones clearly overlap the main locus of blue stars, they provide a less convincing match to the intermediate age turnoffs and corresponding sub-giant populations.  Specifically, the slope of the model SGBs is too flat compared to the data.  The overall agreement between the data and models is much worse than what is observed in some of the more luminous MW satellites, which are known to host genuine intermediate age populations \citep[e.g.,][]{weisz2014a}. Furthermore, the ratio of putative main sequence to SGB stars is unrealistic.  If this was a genuine intermediate age population, we would expect far more MS than SGB stars, given that a star's SGB phase lasts only $\sim10\%$ of its MS lifetime.

In the right hand panel of Figure \ref{fig:bss_cmd}, we over-plot select models of blue stragglers, including their evolution off the main sequence, from \citet{sills2009}.  These models are for a slightly more metal poor population, [M/H]$\sim -2.3$, but \citet{sills2009} suggests that the properties of blue stragglers are not a strong function of metallicity.  As with the intermediate age single stars, the blue straggler models can explain the presence of both the `blue plume' and the scattering of stars across the SGB.  We expect at least some, if not all, of these stars to be blue stragglers, given their ubiquity in other predominantly old stellar systems \cite[e.g.,][]{momany2007, ferraro2015}.  Like the single star models, the blue straggler models: (a) show a slope that is too flat in color relative to the data and (b) do not show the expected ratio of MS to post-MS blue stragglers. However, in relative to the single star models, our understanding of the formation and evolution of blue stragglers is known to be far less complete, which may explain the data-model mismatch.  Specifically, these blue straggler models only reflect collisional products, which can produce a `blue' blue straggler sequence.  The inclusion of binary mass transfer  produces a `red' sequence of blue stragglers \citep[e.g.,][]{lu2010, xin2015}, whose post-MS evolution is less understood.  Thus, it is possible for missing physics to explain the mismatch of blue straggler models, whereas a similar argument is less likely for intermediate age single stars.   Overall, because of our poor understanding of the formation and evolution of blue stragglers, the plotted models should be treated as indicators of the general behavior of blue stragglers.

\section{Discussion}
\label{sec:discussion}

\subsection{Dwarf Galaxy or Globular Cluster?}
\label{sec:caternature}

From our analysis, it is clear that Crater is a GC and not a dwarf galaxy.  Its stellar mass is consistent with being formed at a single age and metallicity, within the precision allowed by a given stellar evolution model.  The sparse population of blue stars above the main turnoff and SGB can plausibly be explained as blue stragglers and their evolved descendants. Moreover, the complete lack of an ancient population ($>$10 Gyr) would make Crater unlike any known dwarf galaxy, all of which are known to host ancient, metal-poor populations \citep[e.g.,][]{tolstoy2009, brown2012, brown2014, weisz2014a}.  Our finding that Crater is a GC is inline with both \citet{laevens2014} and \citet{kirby2015}, who reach the same conclusion using different observations. The strongest evidence in favor of Crater being a dwarf is presented by \citet{bonifacio2015} in which they find a velocity dispersion in excess of expectations from assuming Crater is purely baryonic system.  However, this analysis is based on spectroscopy of only two stars with large uncertainties in their velocities. Using larger samples, \citet{kirby2015} and (M.\ Mateo in prep.) find velocity dispersions that are consistent with expectations for a stellar system comprised entirely of baryons of Crater's mass.  On the whole, the properties of Crater do not satisfy the criteria for being classified as a galaxy as articulated by \citet{willman2012}.

\subsection{An Enigmatic Globular Cluster}
\label{sec:catercluster}

\begin{figure}[t!]
\epsscale{1.2}
\plotone{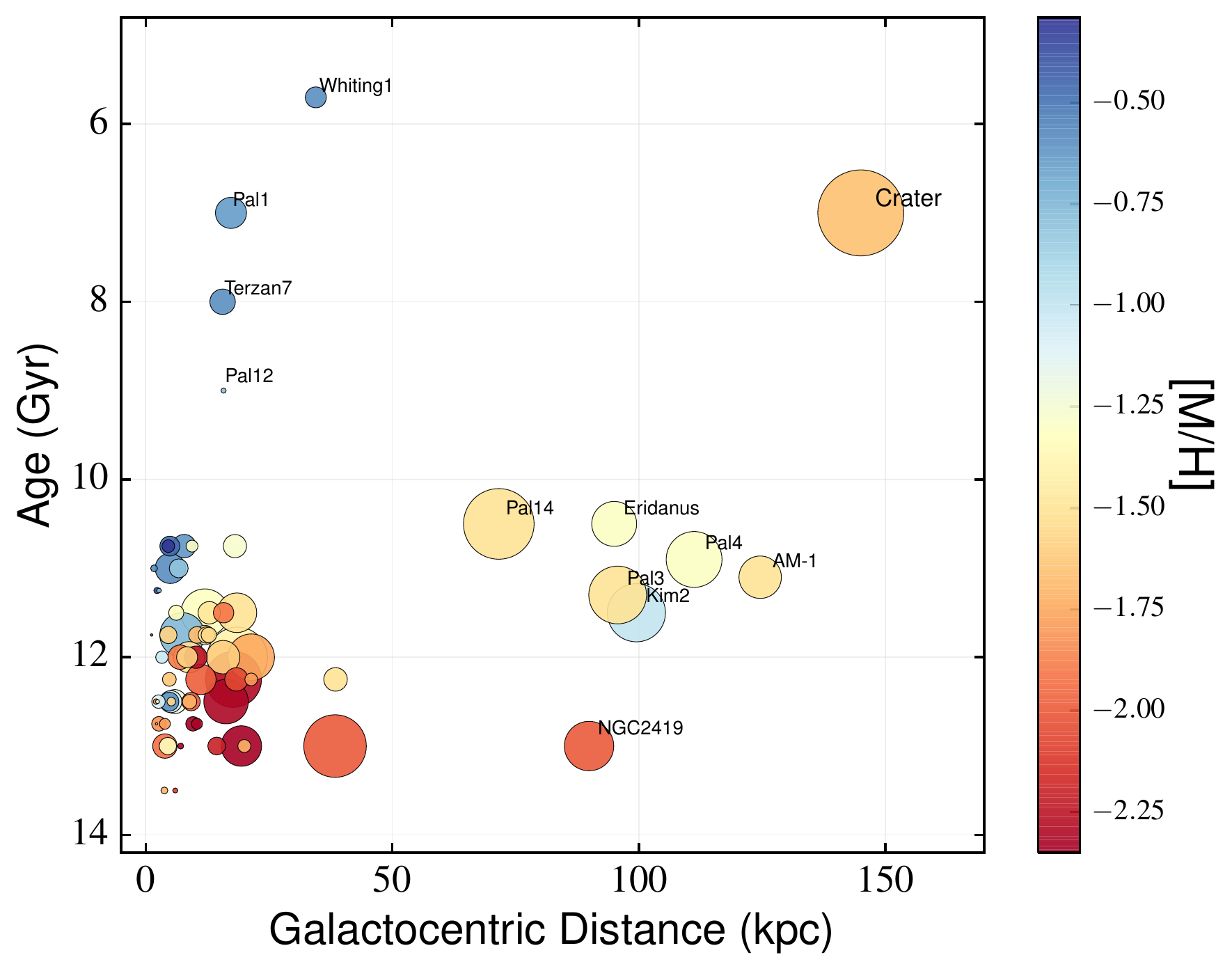}
\caption{The age, metallicity, half-light radius, and galactocentric distance of Crater relative to the general population of MW GCs.  Point sizes are proportional to the cluster half-light radii. Crater is a clear outlier compared to most MW GCs.}
\label{fig:gc_age_dist}
\end{figure}

\begin{figure}[t!]
\epsscale{1.2}
\plotone{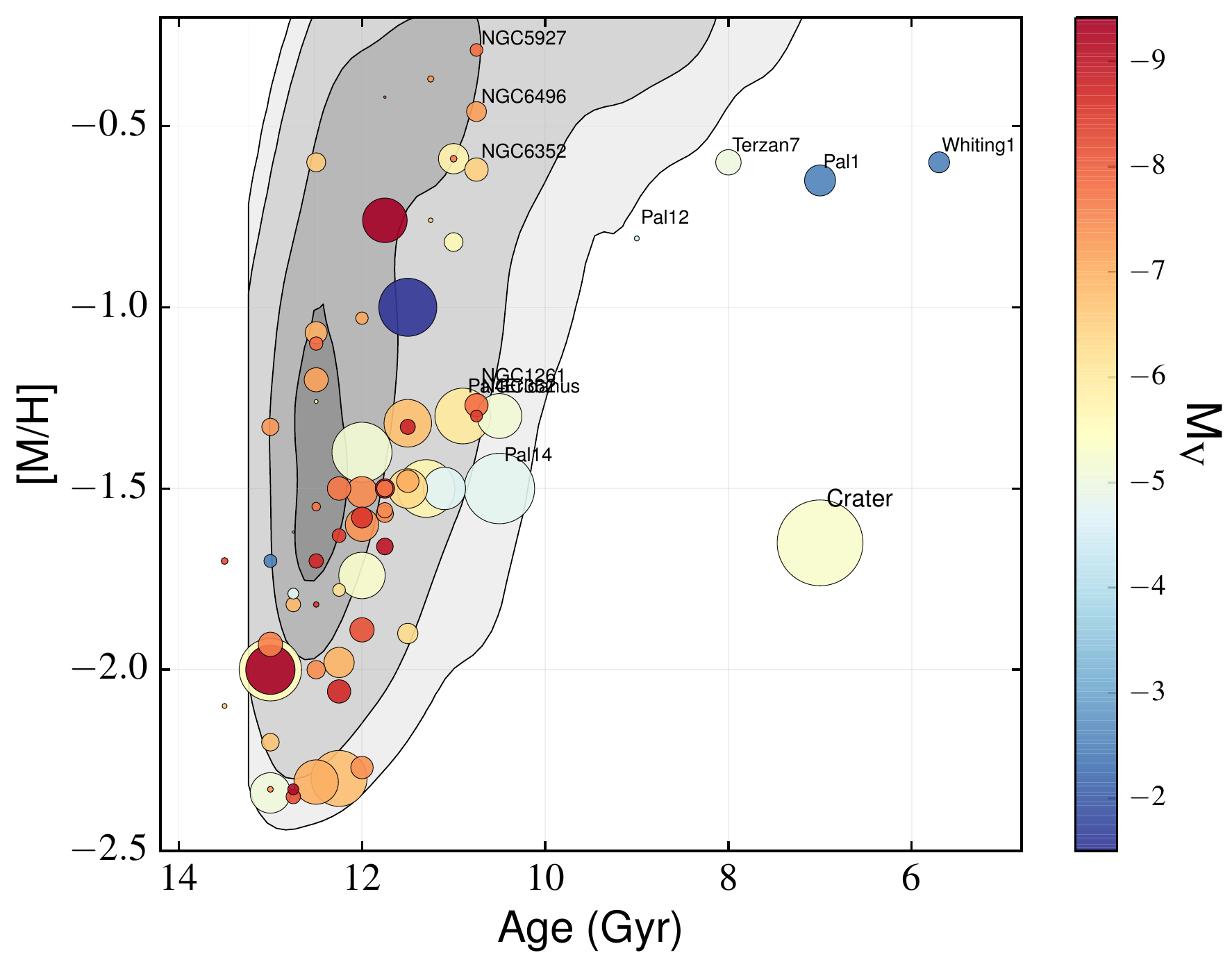}
\caption{The ages and metallicities of MW GCs with predictions from the GC formation models of \citet{li2014} over-plotted as contours.  The point sizes are proportional to the cluster half-light radii.  From darkest to lightest, the contours indicate the expected fraction of GCs: 10\%, 50\%, 90\%, and 97\%.  The models, which posit that GCs form in major mergers, capture the bulk of the MW cluster populations.  However, Crater is a clear outlier, and indicates that mechanisms in addition to mergers are necessary to explain the entire MW GC population.}
\label{fig:gc_lignedin}
\end{figure}

In Figures \ref{fig:gc_age_dist} and \ref{fig:gc_lignedin} we plot Crater's properties relative to those of the general MW GC population. Data for the MW clusters are drawn from a variety of sources, and due to the potential for systematics, we only undertake a qualitative comparison of Crater relative to other clusters.  Specifically, ages and metallicites of the MW GCs are primarily drawn from \citet{vandenberg2013}, where available, and otherwise from \citet{dotter2008} and \citet{dotter2010}.  The distances, luminosities, and sizes are taken from the 2010 update to the MW GC catalog of \citet{harris1996}\footnote{\url{http://www.physics.mcmaster.ca/~harris/mwgc.dat}}.  Properties from newly discovered halo cluster, Kim 2, are from \citet{kim2015b}.  We have not included all known MW GCs on this plot.  For example, we have excluded some of the lowest-lumionsity GCs \citep[e.g., Koposov 1 \&\ 2, Segue 3;][]{koposov2007, belokurov2010, fadely2011}, which  do not have as well-characterized stellar populations.

As shown in Figure \ref{fig:gc_age_dist}, Crater is among the youngest, largest, and most distant of the MW's GCs.  Although it shares a similar metallicity and distance as several halo GCs (e.g., AM-1, Pal 4), its stands out due to its young age, which is comparable only to a handful of GCs located within $\sim$ 40 kpc of the Galactic center (e.g., Whiting 1, Pal 1, Terzan 7).  These young clusters are typically associated with the accretion of Sagittarius; it is clear that Crater is not.

The unusual properties of Crater provide new insight into how various mechanisms (e.g., major merger, accretion of satellites) contributed to the assembly of the MW. To demonstrate this, in Figure \ref{fig:gc_lignedin}, we plot the age-metallicity relationship for the population of MW GCs, including Crater, and  over-plot predictions from the GC formation models presented in \citet{li2014}.  These model combine the dark matter only Millenium {\sc II} simulations \citet{boylankolchin2009} with a semi-analytic analytic model for GC formation as described in \citet{muratov2010}.  These models assume that GCs are formed during major mergers, and are in good qualitative agreement with bulk trends for most MW GCs. 

However, Crater, along with a handful of other young MW GCs, is a significant outlier compared to the model of \citet{li2014}.  This suggests that while major mergers can plausibly explain the bulk of the MW's GC population, other mechanisms (e.g., accretion of low-mass satellite galaxies) are needed to explain these systems.   By virtue of being young, metal-poor, and located in the outer stellar halo, Crater may instead be the signpost of a previously unknown MW accretion event, with its age indicating that such an event occurred more recently than $\sim$ 8 Gyr ago \citep[$z\sim1$, assuming a Planck cosmology;][]{planck2014xvi}.  We discuss the possible origins of Crater further in the following section.

\subsection{Where Did Crater Originate?}
\label{sec:catercluster}

Given its likely extragalactic origin, we can use the properties of Crater to better understand its host galaxy.   Based on its position in the Galaxy, Crater (L$_{\rm MS}$, B$_{\rm MS}$, V$_{GSR}$ $=$ $+$80.6943, $-$5.87460, $+$150\kms; where 'MS' denotes the Magellanic Stream coordinate system) is well-matched to the location of tidal debris from the Magellanic Stream as predicted by the models of \citet{besla2012}.  Furthermore, Crater's heliocentric velocity of $\sim +150$ \kms \citep[]{kirby2015} is consistent with the measured gas velocity of the Magellanic Stream of 100-200 \kms \citep[][]{nidever2010}. Thus, it is plausible that Crater was accreted during an interaction between the MW and the Magellanic Clouds.

\begin{figure}[t!]
\epsscale{1.2}
\plotone{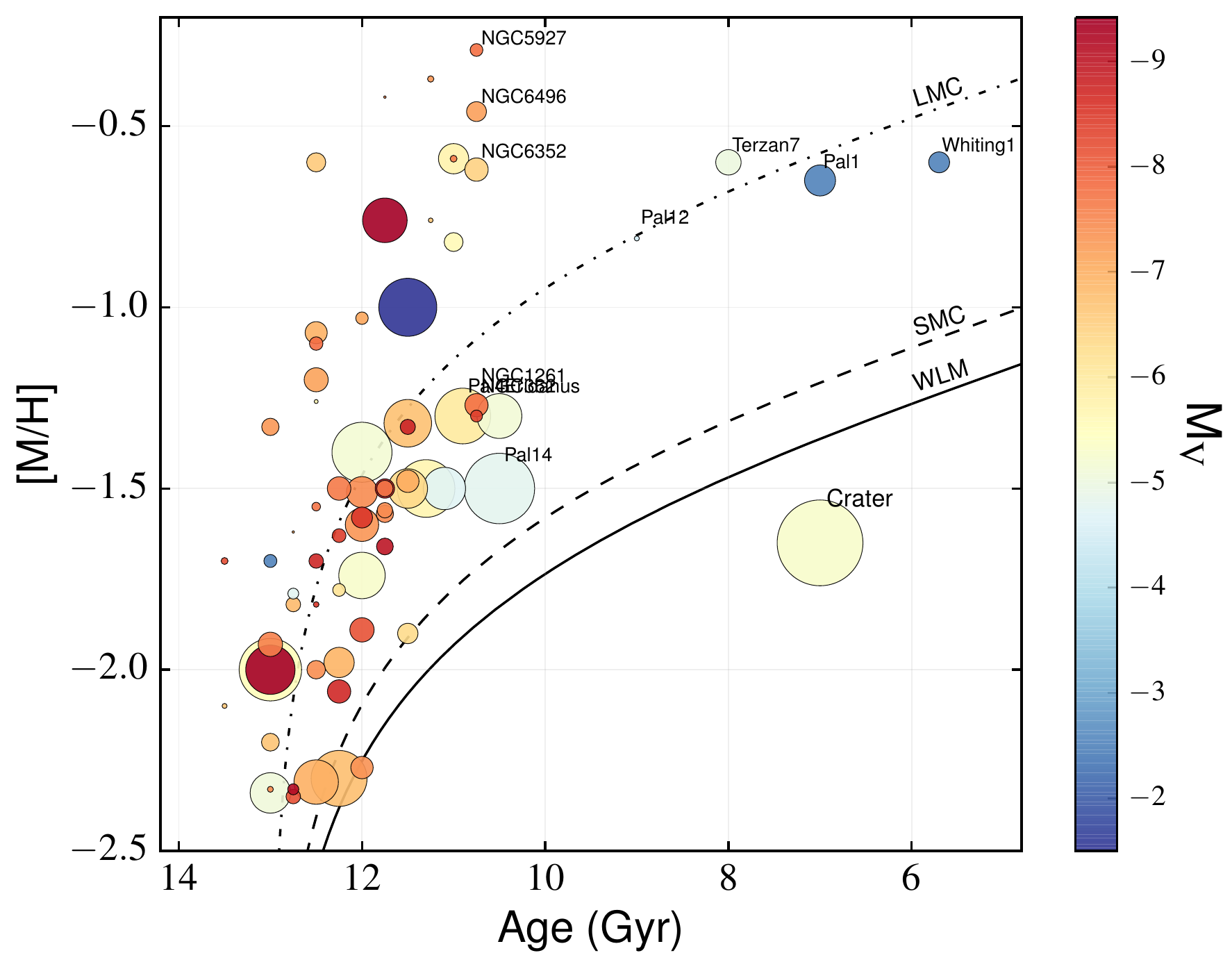}
\caption{The same as Figure \ref{fig:gc_lignedin}, only with the age-metallicity relationships of the LMC, SMC, and WLM as presented in \citet{leaman2013} over-plotted.  Assuming GCs follow their host galaxy's age-metallicity relationship, we suggest that Crater is likely to have formed in a dwarf galaxy less massive than WLM ($M^{z=0}_{star}\sim4\times10^7$ \msun) and possibly in a galaxy similar in mass to Carina or Leo {\sc I}.}
\label{fig:gc_amr}
\end{figure}

However, as shown in Figure \ref{fig:gc_amr}, Crater's age and metallicity are not obviously compatible with the age-metallicity relationships of the LMC or SMC \citep{leaman2013}.  Compared to the SMC age-metallicity relationship, Crater is either too metal poor for its age or too young for its metallicity.  This discrepancy is larger when compared to the LMC.  

It may be the case that Crater is simply an anomalous cluster from the SMC.  A handful of SMC clusters are known to be offest from the galaxy-wide age-metallicity relationship.  A particularly relevant example is that of Lindsay 38, which has an age of $\sim$6.5 Gyr and a metallicity of [M/H]$= -1.49$, as determined from deep HST imaging presented in \citet{glatt2008}.  Its similarity to Crater suggests that is at least plausible that Crater originated as an anomalous cluster in the SMC before being captured by the MW.

Alternatively, if we assume that clusters should roughly follow their host galaxy's age-metallicity relationship, it appears that Crater likely originated in a fairly low-mass dwarf galaxy.  As shown in Figure \ref{fig:gc_amr}, even relative to a less massive dwarf galaxy, WLM \citep[$M^{z=0}_{star}\sim4\times10^7$ \msun;][]{mcconnachie2012}, Crater is slightly offset from the age-metallicity relationship.  Fornax, which is a factor of $\sim$2 less massive than WLM, has a nearly identical age-metallicity relationship.  Although both Fornax and WLM contain GCs, they are general older ($>$10 Gyr) and more metal-poor ([M/H]$<-2.0$) than Crater \citep[e.g.,]{hodge1999, buonanno1998}.  The exception is Fornax-4, which resembles Crater in metallicity, but is 3-4 Gyr older \citep{buonanno1999}.  

Instead, if GCs do trace the age-metallicity relationship of the host, then it is likely that Crater formed in a system similar in mass to Leo {\sc I} or Carina.  Both have had continuous star-formation throughout their lifetimes \citep[e.g.,][]{weisz2014a}, and have present day stellar metallicities similar to Crater.  However, the mean metallicity of Carina, as presented in \citet{deboer2014}, does not reach Crater's value until $\sim$ 5 Gyr ago, indicating it is not an exact match.  Of course, such quantities highly depend on the SFH of a particular system, and this type of mismatch my suggest that Crater's progenitor happened to enrich slightly more quickly than Carina.  On the other hand, there is no known stream near Crater, which decreases the likelihood of Crater's accretion being attributable to a dwarf galaxy that was destroyed by the MW.

\acknowledgments 

The authors thanks Cliff Johnson for interesting discussion about blue stragglers, Ryan Leaman for providing dwarf galaxy age-metallicity relationships, Gurtina Besla for her insight into the Magellanic Stream debris, Jieun Choi for detailed comments, and Oleg Gnedin and Hui Li for providing results from their GC models. DRW is supported by NASA through Hubble Fellowship grant HST-HF-51331.01 awarded by the Space Telescope Science Institute. MGW is supported by National Science Foundation grants AST-1313045 and AST-1412999. EO is supported by NSF grant AST-0807498 and AST-1313006, and MM is supported by NSF grants AST-0808043 and AST-1312997 MG acknowledges the European Research Council (ERC-StG-335936) and the Royal Society for financial support.  Support for this work was provided by NASA through grant number HST-GO-13746.001-A from the Space Telescope Science Institute, which is operated by AURA, Inc., under NASA contract NAS 5-26555. Research leading to these results has also received support from the European Research Council under the European UnionÕs Seventh Framework Program (FP/2007-2013) ERC Grant Agreement no. 308024. Analysis and plots presented in this paper used IPython andpackages from NumPy, SciPy, and Matplotlib \citep[][]{hunter2007,
oliphant2007, perez2007, astropy2013}.

{\it Facility:} \facility{HST (ACS)}

\bibliography{ms.v4.bbl}

\end{document}